\documentclass[10pt,letterpaper,twocolumn]{article} 

\usepackage{ol2}
\usepackage[draft]{hyperref}
\usepackage{amsmath}

\begin{document}

\twocolumn[ 

\title{Optical realization of the dissipative quantum oscillator}


\author{Stefano Longhi and Shane M. Eaton}

\address{Dipartimento di Fisica, Politecnico di Milano and Istituto di Fotonica e Nanotecnologie del Consiglio Nazionale delle Ricerche, Piazza L. da Vinci 32, I-20133 Milano, Italy (stefano.longhi@polimi.it)}

\begin{abstract}
An optical realization of the damped quantum oscillator, based on transverse light dynamics in an optical resonator with slowly-moving mirrors, is theoretically suggested. The optical resonator setting provides a simple implementation of the time-dependent Caldirola-Kanai Hamiltonian of the dissipative quantum oscillator, and enables to visualize the effects of damped oscillations in the classical (ray optics) limit and wave packet collapse in the quantum (wave optics) regime. 
\end{abstract}

\ocis{ (140.4780, 070.5753, 000.1600}
 ] 

Quantum dissipation, i.e. quantum mechanical description of the process of irreversible behavior observed at the classical level, has received a great interest since the earlier developments of quantum physics \cite{r1,r2,r3}. It shares many features with the subjects of quantum decay and quantum decoherence, and hence it is of major importance in different areas of physics. A paradigmatic example of quantum dissipation is provided by the damped quantum oscillator (DQO) \cite{r4,r5}. Two different approaches have been generally considered to describe the DQO. In the first approach \cite{r6,r7,r8,r9} dissipation results from microscopic reversible interaction between the oscillator and the environment. After elimination of the reservoir degrees of freedom, an effective equation of motion that includes a friction force can be obtained \cite{r3,r7}. The other approach incorporates the environment effects by considering an effective time-dependent Hamiltonian \cite{r4,r5}. Such an approach was introduced in two seminal papers by Caldirola and Kanai \cite{r1,r2}. The Caldirola-Kanai Hamiltonian belongs to the class of integrable time-dependent quadratic Hamiltonians, which are of relevance in several areas of physics ranging from quantum optics to cosmology \cite{r10,r11,r12,r13}.\\ In recent years, optics has provided a useful laboratory tool to emulate a wealth of physical phenomena ranging from solid-state physics to relativistic quantum mechanics and relativity \cite{r14,r15,r17bis,r16,r17,r18,r19,r19bis}. In particular,  the use of engineered photonic lattices has been suggested  to emulate quantum decoherence, quantum decay and quantum dissipation within the two above mentioned approaches \cite{r20,r21,r22,r22bis}. For example, light propagation in waveguide arrays with non-nearest neighbor couplings has been shown to map in Fock space the Caldirola-Kanai Hamiltonian \cite{r21}, however special control of waveguide parameters are needed, which is of hard implementation.\par
In this Letter we suggest an optical realization of the Caldirola-Kanai DQO, which is based on transverse light dynamics in an optical resonator with slowly-moving cavity mirrors. The optical cavity setup allows one the visualization of dissipative effects both in the wave-optics (quantum dissipation) and ray-optics (classical dissipation) regimes.\\
The Schr\"{o}dinger equation for the DQO as described by the  Caldirola-Kanai Hamiltonian reads \cite{r1,r2,r4,r5}
\begin{equation}
i \hbar \frac{\partial \psi}{\partial t}= -\frac{\hbar^2}{2m} \exp[-g(t)] \frac{\partial^2 \psi}{\partial x^2}+ \frac{1}{2} m \omega^2 x^2 \exp[g(t)] \psi
\end{equation}
where $g(t)=\gamma t$ for constant friction with damping coefficient $\gamma$. Other forms of $g(t)$, corresponding to a non-exponentially decaying oscillator, can be considered as well \cite{r23}. 
As is well-known, the Caldirola-Kanai Hamiltonian is exactly integrable by means of the 
Lewis-Riesenfeld invariant method or the quantum Arnold transformation \cite{r4,r24}. Indicating by $\phi(X,T)$ the solution to the Schr\"{o}dinger equation for the free-particle, i.e. $i \partial_T \phi=-(\hbar^2 / 2m) \partial^2_X \phi$, the solution to Eq.(1) with the initial condition $\psi(x,0)=\phi(x,0)$ reads explicitly \cite{r24}
\begin{equation}
\psi(x,t) = \frac{1}{\sqrt{u_2(t)}}\exp \left[ \frac{im}{2 \hbar \mathcal{W}(t)} \frac{\dot{u}_2}{u_2} x^2 \right] \phi \left( \frac{x}{u_2(t)}, \frac{u_1(t)}{u_2(t)} \right). 
\end{equation}
In the above equation, the dot indicates the derivative with respect to time $t$, $u_1=u_{1}(t)$ and $u_2=u_2(t)$ are the two linearly-independent solutions to the classical equation of motion
\begin{equation}
\ddot{x}+\dot{g} \dot{x}+\omega^2 u=0
\end{equation}
satisfying the initial conditions $u_1(0)=0$, $\dot u_1(0)=1$, $u_2(0)=1$, $\dot u_2(0)=0$, and $\mathcal{W}(t)=\dot u_1u_2-\dot u_2 u_1= \exp(-g)$ is their Wronskian. For constant friction $\dot{g}= \gamma$, one has $u_1(t)=(1/ \Omega) \exp(- \gamma t /2) \sin ( \Omega t)$ and $u_2(t)=\exp(- \gamma t /2) \cos (\Omega t)+[\gamma / (2 \Omega)] \exp(-\gamma t/2) \sin ( \Omega t)$, where $ \Omega= \sqrt{\omega^2- \gamma^2/4}$. Indicating by $X (T)= \langle \phi | X | \phi \rangle$ and $\Delta X (T)= \langle \phi | (X-X(T))^2 | \phi \rangle^{1/2}$ the mean position and width of the free-particle wave packet $\phi(X,T)$, from Eq.(2) it readily follows that the mean position and width of the DQO are given by $x(t)=u_2(t)X(u_1/u_2)$ and $\Delta x(t)=u_2(t) \Delta X (u_1/u_2)$. From such relations, the following two main properties of the DQO can be stated: (i) the mean position $x(t)$ satisfies the classical equation (3) of the damped oscillator, i.e. the semiclassical motion of the quantum wave packet reproduces the friction of the classical damped oscillator; (ii) since $u_1(t)/ u_2(t)$ is a limited function of time and $u_2(t) \rightarrow 0$ as $t \rightarrow \infty$, one has $\Delta x(t) \rightarrow 0$ as $t \rightarrow \infty$, i.e. the wave packet undergoes a  {\em quantum collapse} in real space and full delocalization in momentum space. In this work we aim to emulate the Caldirola-Kanai Hamiltonian [Eq.(1)] in an optical resonator, highlighting the possibility of a direct visualization of the two above-mentioned properties of the DQO, namely friction motion in the classical (ray optics) limit and wave packet collapse in the quantum (wave optics) regime.\\  
\begin{figure}[htb]
\centerline{\includegraphics[width=8cm]{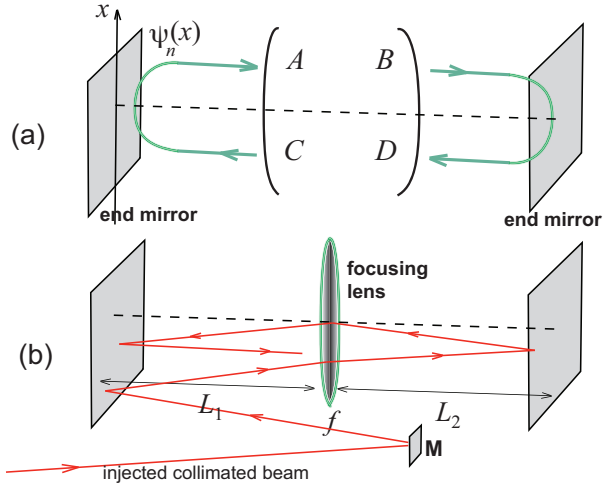}} \caption{ \small
(Color online)  (a) Schematic of wave propagation in a Fabry-Perot resonator in its canonical form representation (flat end mirrors). $ ABCD$ is the resonator round-trip matrix at the end mirror on the left side ($A=D$). (b) Schematic of a resonator that realizes the Caldirola-Kanai DQO. A lens of focal length $f$ is placed inside a plane-plane cavity at distances $L_1$ and $L_2$.  The mirrors are slowly moved at successive round trips to emulate time-dependence in the Schr\"{o}dinger equation (1). In the ray-optics limit, damped harmonic oscillations can be observed by injection of a collimated laser beam at one edge of the resonator, using a small adjustable injection mirror (M), and looking at the beam spots on the left plane mirror at successive bouncing in the cavity.}
\end{figure}
Optical resonators have been deeply investigated in laser physics \cite{r25,r25bis}. Recent theoretical and experimental works have suggested their use to effectively simulate the quantum mechanical Schr\"{o}dinger equation in rather extended forms \cite{r26,r27,r28,r28bis}, for example to emulate a fractional kinetic energy operator \cite{r26,r28} or a gauge (magnetic) field \cite{r27,r28bis}. To mimic the Caldirola-Kanai Hamiltonian, let us consider paraxial beam propagation in a Fabry-Perot optical resonator with spherical optical elements, described by a round-trip matrix $ABCD$ with respect to one of the two end mirrors [Fig.1(a)]. The resonator is assumed to be in its canonical form, i.e. with flat end mirrors, so that \cite{r25bis}
\begin{equation}
A=D= \cos \theta
\end{equation} 
where $\theta$ is real for stability. For the sake of clearness, we consider one transverse spatial  variable $x$, however the analysis holds {\it mutatis mutandis} for a more realistic spherical resonator with two-dimensional transverse coordinates $(x,y)$. Indicating by $\psi_n(x)$ the transverse optical field distribution at the reference flat end mirror in the cavity, beam propagation at successive transits in the cavity is described by the generalized Fresnel integral \cite{r25,r25bis}
\begin{equation}
\psi_{n+1}(x)= \hat{\mathcal{K}} \psi_n(x)=\int_{-\infty}^{\infty} d \xi  \mathcal{K} (x, \xi) \psi_n( \xi). 
\end{equation} 
Neglecting internal cavity losses and output coupling, the kernel $\mathcal{K}$ reads
\begin{equation}
\mathcal{K}(x, \xi)= \sqrt{\frac{i}{\lambda B}} \exp \left[ -\frac{i \pi}{\lambda B} (A \xi^2+Dx^2-2 x \xi) \right]
\end{equation}
where $n$ is the round-trip number and $\lambda$ is the optical wavelength.  For our purposes, it is worth writing Eq.(5) as a stroboscopic map for an associated Schr\"{o}dinger-like wave equation. Using the decomposition $\hat{\mathcal{K}}= \exp (-i \mathcal{H})$ for the generalized Fresnel integral operator, where $\hat{\mathcal{H}}$ is a second-order linear differential operator \cite{r29,r30}, the map (5) $\psi_{n+1}=\exp(-i \hat{\mathcal{H}}) \psi_n$ can be derived from the Schr\"{o}dinger equation $i \partial_n \psi(x,n)= \hat{\mathcal{H}} \psi(x,n)$, which reads explicitly
\begin{equation}
i \frac{\partial \psi}{\partial n}= \frac{B \theta}{2 k \sin \theta} \frac{\partial^2 \psi}{\partial x^2}+ \frac{k \theta C}{2 \sin \theta} x^2 \psi,
\end{equation}
where $k= 2 \pi / \lambda$ is the optical wave number and the angle $\theta$ is defined by Eq.(4). Let us now suppose that the $ABCD$ resonator matrix elements are slowly varied at successive round trips, in such a way that $A=D= \cos \theta$ is kept constant whereas 
\begin{equation}
B(n)=B(0)\exp[-g(n)], \; \; C(n)=C(0) \exp[g(n)]
\end{equation} 
where $g(n)$ is a slowly-increasing function of $n$ with $g(0)=1$. As discussed below, in practice this can be simply achieved by slowly moving the end mirrors of the resonator. 
\begin{figure}[htb]
\centerline{\includegraphics[width=8.2cm]{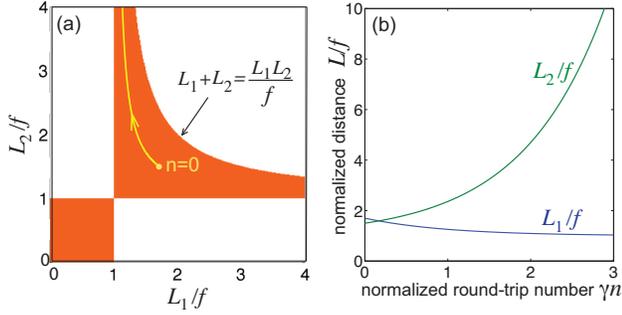}} \caption{ \small
(Color online)  (a) Stability diagram of the resonator of Fig.1(b) in the $(L_1/f, L_2/f)$ plane. The stability region comprises two domains, depicted by the shaded areas. (b) Behavior of normalized mirror distances $L_{1}/f$ and $L_2/f$ versus normalized round-trip number $\gamma n$ that realizes a constant friction $g(n)=\gamma n$. The initial mirror distances are $L_1(0)/f=1.7$ and $L_2(0)/f=1.5$, corresponding to an angle $\theta \simeq 1.88$ rad. The continuous curve inside the upper stability domain in panel (a) shows the path followed by  $(L_1/f,L_2/f)$ as the round trip number increases from $n=0$.}
\end{figure}
After substitution of Eq.(8) into Eq.(7), one readily recognizes that Eq.(7) is analogous to the Caldirola-Kanai  Schr\"{o}dinger equation (1), provided that the following formal substitutions are made
\begin{subequations}
\begin{eqnarray}
t & \rightarrow & n \; , \;\;\;
\omega  \rightarrow  \theta  \\
\hbar & \rightarrow & 1/k \; , \;\;\;
m  \rightarrow  \frac{1}{\theta} \sqrt{-\frac{C(0)}{B(0)}}.
\end{eqnarray}
\end{subequations}
Therefore, wave propagation inside the optical cavity at successive round trips with slowly-varying elements $B$ and $C$ emulates the Schr\"{o}dingier equation (1) of the DQO. For constant friction $g(n)=\gamma n$, the requirement that $B$ and $C$ vary slowly over one round trip time implies $\gamma \ll 1$. For a resonator well inside the stability domain, i.e. $\theta$ far from enough from zero, from Eq.(9a) one has $\gamma / \omega \ll 1$, i.e. our optical system emulates the DQO in the {\em underdamped} regime. To reach critical damping, the resonator must operate close to the instability region $\theta \simeq 0$, which might be challenging in an experiment.\\
To obtain the slow changes of  matrix elements $B$ and $C$ as in Eq.(8), let us consider as an example the cavity shown in Fig.1(b), which comprises two flat end mirrors and one focusing lens (focal length $f$) placed at distances $L_1$ and $L_2$ from the two mirrors. The stability region of the resonator is depicted in Fig.2(a). A slow change of parameters $B$ and $C$ can be obtained by slowly moving the two end mirrors, i.e. by making the distances $L_1(n)$ and $L_2(n)$ slowly varying with the round trip number $n$ with an initial condition $L_1(0)$ and $L_2(0)$ in the upper stability domain of Fig.2(a). The behavior of $L_1(n)$ and $L_2(n)$ that realizes the damping $g(n)$ is obtained as a solution of the equation
\begin{equation}
\left(
\begin{array}{c}
dL_1 \\
dL_2
\end{array}
\right)=
\left(
\begin{array}{cc}
\frac{\partial B}{\partial L_1} & \frac{\partial B}{\partial L_2} \\
\frac{\partial C}{\partial L_1} & \frac{\partial C}{ \partial L_2}
\end{array}
\right)^{-1}
\left(
\begin{array}{c}
-B \\
C
\end{array}
\right) \left( \frac{dg}{dn} \right)  dn
\end{equation}
with some given initial conditions $L_1(0)$ and $L_2(0)$, where
\begin{eqnarray}
B & = & 2 \left( 1-\frac{L_1}{f} \right) \left(L_1+L_2-\frac{L_1L_2}{f} \right) \;\;\; \\
C & = & -\frac{2}{f}  \left(1-\frac{L_2}{f} \right) 
\end{eqnarray}
are the $B$ and $C$ matrix elements. After integration one obtains
\begin{eqnarray}
L_2(n) & = & f+[L_2(0)-f] \exp[g(n)] \\
L_1(n) & = & f \frac{2L_2(n)-f(1-\cos \theta)}{2L_2(n)-2f}.
\end{eqnarray}
As an example, Fig.2(b) shows the paths of $L_1$ and $L_2$ that realize an exponential damping $g(n)= \gamma n$. Note that, since $\theta$ does not change when $L_1$ and $L_2$ are varied, the resonator remains inside the stability domain [solid curve in Fig.2(a)].\\ 
 The optical cavity setup enables in a simple way to visualize the two main effects of the DQO, namely (i) the damped oscillatory trajectory of the wave packet, and (ii) its quantum collapse. To observe the damped motion of the oscillator, it is worth considering the ray optics limit of beam propagation inside the resonator. In the ray optics limit, damped harmonic oscillations sustained by the optical cavity with slowly-moving mirrors can be observed by injection of a collimated laser beam (e.g. an He-Ne laser) at one edge of the resonator and looking at the motion of beam spots on the left plane mirror at successive round trips [Fig.1(b); see also the setup discussed in Fig.15.16 of Ref.\cite{r25}). According to paraxial ray matrix analysis, the ray displacement $x_n$ and ray angle $x'_n$ of the beam spot on the flat mirror at the $n$-th round trip satisfy the recurrence relation \cite{r25bis}
 \begin{equation}
 \left(
 \begin{array}{c}
 x_{n} \\
 x'_{n}
 \end{array}
 \right)=
 \left(
 \begin{array}{cc}
 A(n) & B(n) \\
 C(n) & D(n)
 \end{array}
 \right) \times 
 \left(
 \begin{array}{c}
 x_{n-1} \\
 x'_{n-1}
 \end{array}
 \right).
 \end{equation}
 From Eq.(15) the following second-order difference equation for $r_n$ is readily obtained
 \begin{equation}
 x_{n+1}+\frac{B(n+1)}{B(n)}x_{n-1}- \left[ A(n+1) + \frac{B(n+1)D(n)}{B(n)} \right] x_n=0.
 \end{equation}
  In our case $A(n)=D(n)= \cos \theta$ is independent of $n$, and for a constant friction, $g(n)=\gamma n$, one has $B(n+1)/B(n)=\exp(-\gamma)$. Hence from Eq.(16) one obtains
 \begin{equation}
 x_{n+1}- \cos \theta [1+\exp(-\gamma)] x_{n}+ \exp(-\gamma) x_{n-1}=0.
 \end{equation} 
 The solution to Eq.(17) is given by  $x_n=\alpha \mu_1^n+ \beta \mu_2^n$, where $\mu_{1,2}$ are the roots of the algebraic equation $\mu^2-\cos \theta [1+\exp(-\gamma)] \mu+ \exp(-\gamma)=0$ and $\alpha$, $\beta$ are determined by the initial condition $x_0$, $x'_0$. Note that $|\mu_{1,2}|= \exp(-\gamma)$, and that $\mu_{1,2} \simeq \exp(-\gamma \pm i \theta)$ for $\gamma \ll 1$. Therefore, as expected  $x_n$ describes an exponentially-damped oscillation with period $2 \pi / \theta$ and damping parameter $\gamma$. An example of the behavior of $x_n$ for the moving cavity mirror parameters of Fig.2 is shown in Fig.3(a). It should be noted that in a realistic optical resonator with two transverse spatial dimensions $(x,y)$ the beam spots at coordinates $(x_n,y_n)$ are distributed along a general elliptic curve in the transverse plane, i.e. they describe Lissajous patterns typical of the two-dimensional harmonic oscillator (see, for example, Ref.\cite{r25}, p. 603). In the cavity with mirrors at rest the Lissajous pattern is stationary, whereas when the two mirrors are synchronously moved following the laws (13) and (14) the Lissajous pattern shrinks toward the optical axis $x=y=0$; see Fig.3(b). In an experiment, the maximum damping rate $\gamma$ (i.e. rate of Lissajous pattern shrinking) that one can achieve is ultimately limited by the finite speed at which mechanical mirrors can be moved. To estimate realistic values of $\gamma$, let us calculate the speed $v$ of faster moving mirror, i.e. of mirror 2. From Eq.(13) one has $v=|dL_2/dt|=(\gamma / T_R)[L_2(t)-f]$, where $T_R=(L_1+L_2)/c$ is the round-trip transit of photons in the resonator and $t=nT_R$. Taking $L_1 \simeq f$ and $L_2 \gg f$, which holds after some transient at initial round trips [see solid curve in Fig.2(a)], one has $v \simeq \gamma c$.  Values of $v$ achievable with motorized linear stage guides with travel up to $ \sim 1$m are $v \sim 300-1000 \; {\rm mm/s}$, which yields $\gamma \sim v/c \sim 10^{-8}-10^{-9}$. At such small damping rates, the dynamics in the resonator is fully adiabatic, so that in an experiment one can simply record Lissajous patterns at fixed mirror positions and then move them following the laws (13) and (14).

 \begin{figure}[htb]
\centerline{\includegraphics[width=8.2cm]{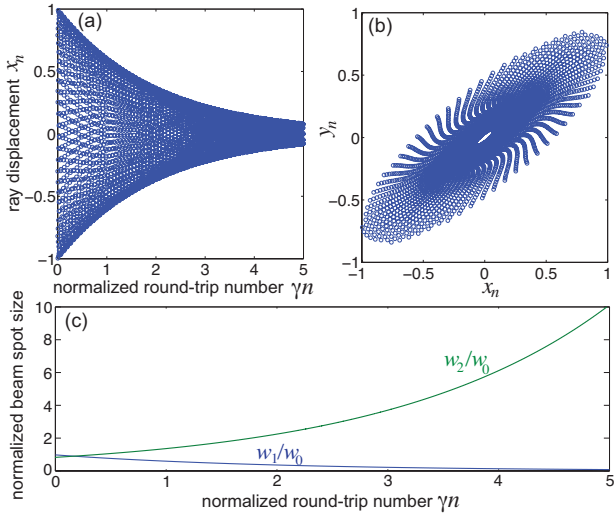}} \caption{ \small
(Color online)  (a) Behavior of the ray displacement $x_n$ at successive round trips in the cavity for $x_0=1$, $x'_0=0$, $\gamma=10^{-3}$  and for  the same parameter values as in Fig.2(b) [$L_1(0)/f=1.7$ and $L_2(0)/f=1.5$, corresponding to an angle $\theta \simeq 1.88$ rad]. (b) Example of a contracting Lissajous pattern in the two-dimensional transverse plane for initial condition $x_0=1$, $x'_0=0$, $y_0=0.7$, $y_0'=0.5$. Lower values of damping $\gamma$ just result in more dense spots. (c) Behavior of beam spot sizes $w_1$ and $w_2$ of the TEM$_{\rm 00}$ Gaussian mode on the two flat mirrors, normalized to $w_0= \sqrt{\lambda f / \pi}$, versus normalized round-trip number $\gamma n$.}
\end{figure}
The ray optics analysis and Lissajous patterns discussed above capture the classical limit of the DQO, i.e. damped classical oscillator dynamics.  Genuine quantum effects of the DQO with no classical counterpart, which have been studied in previous works, can be emulated by considering wave-optics dynamics in the resonator. Such effects include existence of pseudo-stationary states in the underdamped regime \cite{r5}, dissipation-induced transitions \cite{referee1} and wave packet spreading suppression by dissipation \cite{referee2}. Interestingly, pseudo-stationary states correspond to a rest particle but with a wave function that eventually undergoes a wave collapse. Wave collapse can be visualized by considering an active cavity, i.e. a laser oscillator with the optical resonator of Fig.1(b). 
The laser is operated above threshold in continuous-wave mode and transverse mode selection in the fundamental TEM$_{\rm 00}$ mode is accomplished by e.g. longitudinal pumping \cite{r25bis}.
The transverse distribution of the output laser beam can be monitored on a CCD camera at fixed mirror positions, and then moving the mirrors as discussed above. When the flat end mirrors of the resonator are slowly moved, the beam spot size $w_1(n)$ at the output left mirror undergoes adiabatic changes according to the relation
 \begin{equation}
 w_1(n)=w_1(0) \sqrt{u_2^2(n)+\theta^2 u_1^2(n)} \simeq w_1(0) \exp(- \gamma n /2)
 \end{equation}  
  where $w_1(0)=(\lambda / \pi)^{1/2} [-B(0)/C(0)]^{1/4}$ is the spot size of the initial TEM$_{\rm 00}$ mode. The shrinking of the beam spot size at successive round trips, $w_1(n) \rightarrow 0$ as $n \rightarrow \infty$,  is a clear signature of wave packet collapse. In our optical setup, wave packet collapse arises because the resonator stability region shrinks as $L_2/f \rightarrow \infty$ and $L_1/f \rightarrow 1^+$, i.e. a stability boundary is reached even though the angle $\theta$ does not change [see the solid curve in Fig.2(a)]. In such a limit, the flat mirror at the left side of the resonator gets close to the focal lens plane, whereas the other one is its Fourier-conjugate plane. While the spot size $w_1(n)$ on the left mirror shrinks as $n$ increases, the corresponding beam spot size $w_2(n)$ on the right flat mirror enlarges; see Fig.3(c). Interestingly, the product $w_1 w_2$ is independent of $n$ and reads 
 \begin{equation}
 w_1(n) w_2(n) = \frac{\lambda f}{\pi} \sqrt{\frac{1- \cos \theta}{2}},
 \end{equation} 
which emulates the position-momentum uncertainty relation.\par
In conclusion, we suggested a simple optical realization of the Caldirola-Kanai Hamiltonian describing a damped quantum harmonic oscillator, which is based on the transverse ray and wave dynamics of light in an optical cavity with slowly-moving mirrors. As compared to recent proposals of Caldirola-Kanai Hamiltonians using specially-tailored photonic lattices with non-nearest neighbor couplings and longitudinal waveguide modulation \cite{r21,r22}, the optical resonator setup proposed in this work enables a much simpler implementation as well as a direct visualization of wave packet dynamics, such as damped Lissajous trajectories and wave packet collapse. Our results also indicate that transverse light dynamics in optical resonators with moving mirrors provides an accessible testbed for emulating in optics quadratic time-dependent Hamiltonians, and thus it can be of potential interest for the simulation on a tabletop of other physical phenomena or exotic theoretical models based on time-dependent quadratic Hamiltonians, such as the behavior of gravitational waves  in a Friedmann-Robertson-Walker spacetime \cite{r13,r31} or gravitational decoherence in non-linear dispersion cosmological models \cite{r32}. \\
\\                                                                                                                                                                                                                                                                                                                                                                                                                                                                                                                                                                                                                                                                                                                                                                                                                                                                                                                                                                                                                                                                                                                                                                                                                                                                                                                                                                                                                                                                                                                                                                                                                                                                                                                                                                                                                                                                                                                                                                                                                                                                                                                                                                                                                                                                                                                                                                                                                                                                                                                                                                                                                                                                                                                                                                                                                                                                                                                                                                                                                                                                                                                                                                                                                                                                                                                                                                                                                                                                                                                                                                                                                                                                                                                                                                                                                                                                                                                                                                                                                                                                                                                                                                                                                                                                                                                                                                                                                                                                                                                                                                                                                                                                                                                                                                                                                                                                
\noindent
 Cariplo ERC support project "FemtoDiamante" is gratefully acknowledged. The article is dedicated to Professor Orazio Svelto on the occasion of his 80th birthday.

\newpage

\footnotesize {\bf References with full titles}\\
\\
\noindent
1. P. Caldirola, "Forze non conservative nella meccanica quantistica", Nuovo Cim. {\bf 18}, 393 (1941).\\
2. E. Kanai, "On the Quantization of the Dissipative Systems", Prog. Theor. Phys. {\bf 3}, 440 (1948).\\ 
3. U. Weiss, {\it Quantum dissipative systems} (World Scientific, Singapore, 2008, 3rd ed.). \\
4. H. Dekker, "Classical and quantum mechanics of the damped harmonic oscillator", Phys. Rep. {\bf 80}, 1 (1981).\\
5. C.-I. Um, K.-H. Yeon, and T.F. George, "The quantum damped harmonic oscillator", Phys. Rep. {\bf 362}, 63 (2002).\\
6. R.P. Feynman and F.L. Vernon, "The theory of a general quantum system interacting with a linear dissipative system", Ann. Phys. {\bf 24}, 
118 (1963).\\
7. P. Ullersma, "An exactly solvable model for Brownian motion : I. Derivation of the Langevin equation", Physica {\bf 32},  27 (1966).\\
8. O.A. Caldeira and  A.J. Leggett, "Influence of Dissipation on Quantum Tunneling in Macroscopic Systems", Phys. Rev. Lett. {\bf 46}, 211 (1981).\\
9. O.A. Caldeira and  A.J. Leggett, "Quantum tunnelling in a dissipative system", Ann. Phys.  {\bf 149}, 374 (1983).\\
10. C. K. Law, "Effective Hamiltonian for the radiation in a cavity with a moving mirror and a time-varying dielectric medium," Phys. Rev. A {\bf 49}, 433 (1994).\\
11. I. A. Pedrosa, C. Furtado, and A. Rosas, "Light propagation: from dielectrics to curved spacetimes," EPL {\bf 94}, 30002 (2011).\\
12. N.C. Menicucci and G.J. Milburn, "Single trapped ion as a time-dependent harmonic oscillator," Phys. Rev. A {\bf 76}, 052105 (2007).\\
13. A. Geralico, G. Landolfi, G. Ruggeri, and G. Soliani, "Novel approach to the study of quantum effects in the early universe," Phys. Rev. D {\bf 69}, 043504 (2004).\\
14. S. Longhi, "Quantum-optical analogies using photonic structures", Laser \& Photon. Rev. {\bf 3}, 243 (2009).\\
15. S. Longhi, "Classical simulation of relativistic quantum mechanics in periodic optical structures", Appl. Phys. B {\bf 104}, 453 (2011).\\
16. T.G. Philbin, C. Kuklewicz, S. Robertson, S. Hill, F. Konig, and U. Leonhardt, "Fiber-optical analogue of the event horizon", Science {\bf 319}, 1367 (2008).\\
17. S. Longhi, "Klein tunneling in binary photonic superlattices", Phys. Rev. B {\bf 81}, 075102 (2010).\\
18. V.H. Schultheiss, S. Batz, A. Szameit, F. Dreisow, S. Nolte, A. T\"{u}nnermann, S. Longhi, and U. Peschel, "Optics in curved space", Phys. Rev. Lett. {\bf 105}, 143901 (2010).\\
19. M.C. Rechtsman, J.M. Zeuner, A T\"{u}nnermann, S Nolte, M Segev, and A Szameit, "Strain-induced pseudomagnetic field and photonic Landau levels in dielectric structures", Nature Photon. {\bf 7}, 153 (2013).\\
20. S. Gentilini, M.C. Braidotti, G. Marcucci, E. DelRe, and C. Conti, "Physical realization of the Glauber quantum oscillator", Sci. Rep. {\bf 5}, 15816 (2015).\\
21. R. Bekenstein,	R. Schley,	M. Mutzafi, C. Rotschild, and M. Segev, "Optical simulations of gravitational effects in the Newton?Schr\"{o}dinger system", Nature Phys. {\bf 11}, 872 (2015).\\
22. S. Longhi, "Quantum simulation of decoherence in optical waveguide lattices", Opt. Lett. {\bf  38}, 4884 (2013).\\
23. B. M. Rodriguez Lara, P. Aleahmad, H.M. Moya-Cessa, and D.N. Christodoulides, "Ermakov-Lewis symmetry in photonic lattices", Opt. Lett. {\bf 39}, 2083 (2014).\\
24. D. Stefanatos, "Design of a photonic lattice using shortcuts to adiabaticity", Phys. Rev. A {\bf 90}, 023811 (2014).\\
25. A. Crespi, L. Sansoni, G. Della Valle, A. Ciamei, R. Ramponi, F. Sciarrino, P. Mataloni, S. Longhi, and R. Osellame, "Particle Statistics Affects Quantum Decay and Fano Interference", Phys. Rev. Lett. {\bf 114}, 090201 (2015).\\
26. J.R. Choi, "The effects of nonextensivity on quantum dissipation", Sci. Rep. {\bf 4}, 3911 (2014).\\
27. V. Aldaya, F. Cossio, J. Guerrero, and F.F. Lopez-Ruiz, "The quantum Arnold transformation," J. Phys. A {\bf 44}, 065302 (2011).\\ 
28. A.E. Siegman, {\it Lasers} (University Science Books, Mill Valley, CA, 1986).\\
29. O. Svelto, {\it Principles of Lasers} (Springer, New York, 2010, fifth ed.), chaps. 4 and 5.\\ 
30. S. Longhi, "Fractional Schr\"{o}dinger equation in optics", Opt. Lett. {\bf 40}, 1117 (2015).\\ 
31. S. Longhi, "Synthetic gauge fields for light beams in optical resonators", Opt. Lett. {\bf 40}, 2941 (2015).\\
32. Y. Zhang, X. Liu, M.R. Belic, W. Zhong, Y. Zhang, and M. Xiao, "Propagation Dynamics of a Light Beam in a Fractional Schr\"{o}dinger Equation", Phys. Rev. Lett. {\bf 115}, 180403 (2015).\\
33. N. Schine, A. Ryou, A. Gromov, A. Sommer, and J. Simon, "Synthetic Landau levels for photons", arXiv:1511.07381 (2015).\\
34. A.M. Dunlop, W.J. Firth, D.R. Heatley, and E.M. Wright, "Generalized mean-field or master equation for nonlinear cavities with transverse effects", Opt. Lett. {\bf 21}, 770 (1996).\\
35. S. Longhi, "Extended matrix method for Gaussian pulse propagation and generalized mode-locking master equation", Opt. Commun. {\bf 188}, 239 (2001).\\
36. P. Croxson, {\it Induced transitions and energy of a damped oscillator}, Phys. Rev. A {\bf 49}, 588 (1994).\\
37.C.-P. Sun and L.-H. Yu, {\it Exact dynamics of a quantum dissipative system in a constant external field}, Phys. Rev. A {\bf 51}, 1845 (1995).\\
38. I.A. Pedrosa , C. Furtado, and A. Rosas, "Exact linear invariants and quantum effects in the early universe", Phys. Lett. B {\bf 651}, 384 (2007).\\
39. G. Landolfi and G. Soliani, "Trans-Planckian effects in nonlinear-dispersion cosmologies", Phys. Lett.B {\bf 588}, 1 (2004). 

\end{document}